\documentclass[runningheads]{llncs}
\usepackage[T1]{fontenc}
\usepackage{graphicx,verbatim}
\usepackage{array}
\usepackage{tcolorbox}
\usepackage{multirow}
\usepackage{enumitem}
\usepackage{hyperref}
\usepackage{float}
\begin{document}
\title{Exploring the Design Space of 3D MLLMs for CT Report Generation}
%

\author{Mohammed Baharoon\inst{1, 2*} \and
Jun Ma\inst{1, 7*} \and
Congyu Fang\inst{1, 3, 5} \and
Augustin Toma\inst{1, 4} \and \ \
Bo Wang\inst{1, 3, 5, 6, 7\textsuperscript{\textdagger}}
}

\institute{Vector Institute for Artificial Intelligence, Toronto, Canada \and Department of Biomedical Informatics, Harvard Medical School, Boston, Massachusetts \and Peter Munk Cardiac Centre, University Health Network, Toronto, Canada \and
Medical Biophysics, University of Toronto, Toronto, Canada \and
Department of Computer Science, University of Toronto, Toronto, Canada \and
Department of Laboratory Medicine and Pathobiology, University of Toronto,  Toronto, Canada \and
AI Hub, University Health Network, Toronto, Canada \\
\* * Equal contribution \\
\textsuperscript{\textdagger} Corresponding Author \\
}

\maketitle              
\begin{abstract}
Multimodal Large Language Models (MLLMs) have emerged as a promising way to automate Radiology Report Generation (RRG). In this work, we systematically investigate the design space of 3D MLLMs, including visual input representation, projectors, Large Language Models (LLMs), and fine-tuning techniques for 3D CT report generation. We also introduce two knowledge-based report augmentation methods that improve performance on the GREEN score by up to 10\%, achieving the 2nd place on the MICCAI 2024 AMOS-MM challenge. Our results on the 1,687 cases from the AMOS-MM dataset show that RRG is largely independent of the size of the LLM under the same training protocol. We also show that larger volume size does not always improve performance if the original ViT was pre-trained on a smaller volume size. Lastly, we show that using a segmentation mask along with the CT volume improves performance. The code is publicly available at \href{https://github.com/bowang-lab/AMOS-MM-Solution}{https://github.com/bowang-lab/AMOS-MM-Solution}. 


\end{abstract}
\section{Introduction}

Computed Tomography (CT) is a cornerstone of modern diagnostic imaging, offering detailed insights into internal anatomical structures and playing a critical role in diagnosing a wide range of diseases \cite{ct}. However, the rapid increase in the need for CT examinations presents a significant challenge for radiologists, who must interpret complex 3D volumetric data and generate comprehensive reports under tight time constraints \cite{rise_ct}. This growing demand places immense pressure on healthcare systems, often leading to delays in diagnosis and treatment, which can adversely affect patient outcomes \cite{work_load}.

To address this challenge, there has been growing interest in developing automated systems using Multimodal Large Language Models (MLLMs) for radiology report generation~\cite{medversa,maira,m3d,radfm,ct2rep}, leveraging their ability to process medical images with advanced natural language generation. A representative example is the Large Language and Vision Assistant (LLaVA)~\cite{llava}, where a vision encoder processes input images, and a projector transforms the encoded features into the language embedding space. These projected embeddings are concatenated with natural language instructions and fed into a language decoder to generate text responses conditioned on both the image and instruction embeddings. LLaVA has been extensively explored for the generation of radiology reports from 2D images~\cite{maira,cxr-llava,llava-med}, with adaptations for the medical domain. For example, MAIRA-2 uses the LLaVA framework for the generation of X-ray reports with localized findings, resulting in more grounded reports \cite{maira}. 

Recent research has expanded to MLLMs for 3D medical images, which offer richer spatial information but introduce computational challenges such as handling volumetric data and managing the high token count~\cite{ct2rep,radfm,m3d,medversa}. Works, such as M3D~\cite{m3d} and RadFM~\cite{radfm}, adopt the LLaVA framework by using a vision-to-language embedding projector. To address the high dimensionality of 3D image embeddings, M3D introduces a spatial pooling layer, while RadFM employs a perceiver module~\cite{radfm}.

In this study, we systematically investigate the design space of 3D MLLMs for radiology report generation from CT scans. Our work explores key architectural choices, including visual representation, projectors, LLMs, and fine-tuning strategies. We also introduce heuristic report augmentation methods to improve the completeness of generated reports. The main contributions are summarized as follows: 
\begin{itemize}[label=\textbullet]
    \item \textbf{Decoupled architecture design:} we decompose MLLMs into plug-and-play modules, allowing for comprehensive studies to obtain better insights on how these components contribute to performance. 
    \item \textbf{Knowledge-based argumentation:} we introduce two knowledge-based report augmentation methods to enhance report completeness, increasing the performance by over 10\%, from 0.366 to 0.470, on the GREEN metric.
    \item  \textbf{State-of-the-art performance:} our solution achieved the second place in the hidden test set of the MICCAI 2024 AMOS-MM challenge.
    \item \textbf{Modular implementation:} all methods are implemented and open-sourced in a unified framework, ensuring reproducibility, fair comparisons, and easy integration of new methods.
\end{itemize}

\section{Methods}

In this work, we focus on MLLMs that follow the LLaVA architecture because of their simple yet effective and popular design \cite{llava}, which includes an image encoder, a projector, and an LLM. Specifically, a Vision Transformer (ViT) \cite{ViT} is used to extract visual embeddings from an image, followed by a projector to map these embeddings from the image space to the LLM input space, which are then passed to the LLM as input, along with the query prompt. Next, in Section \ref{sec:projector}, and Section \ref{sec:llm} we explore different design choices for each element in the MLLM. Lastly, in Section \ref{sec:report_completion}, we introduce our knowledge-based report augmentation methods, which help to ensure that the final reports are comprehensive.

\begin{figure}[h]
    \centering
    \includegraphics[width=1\textwidth]{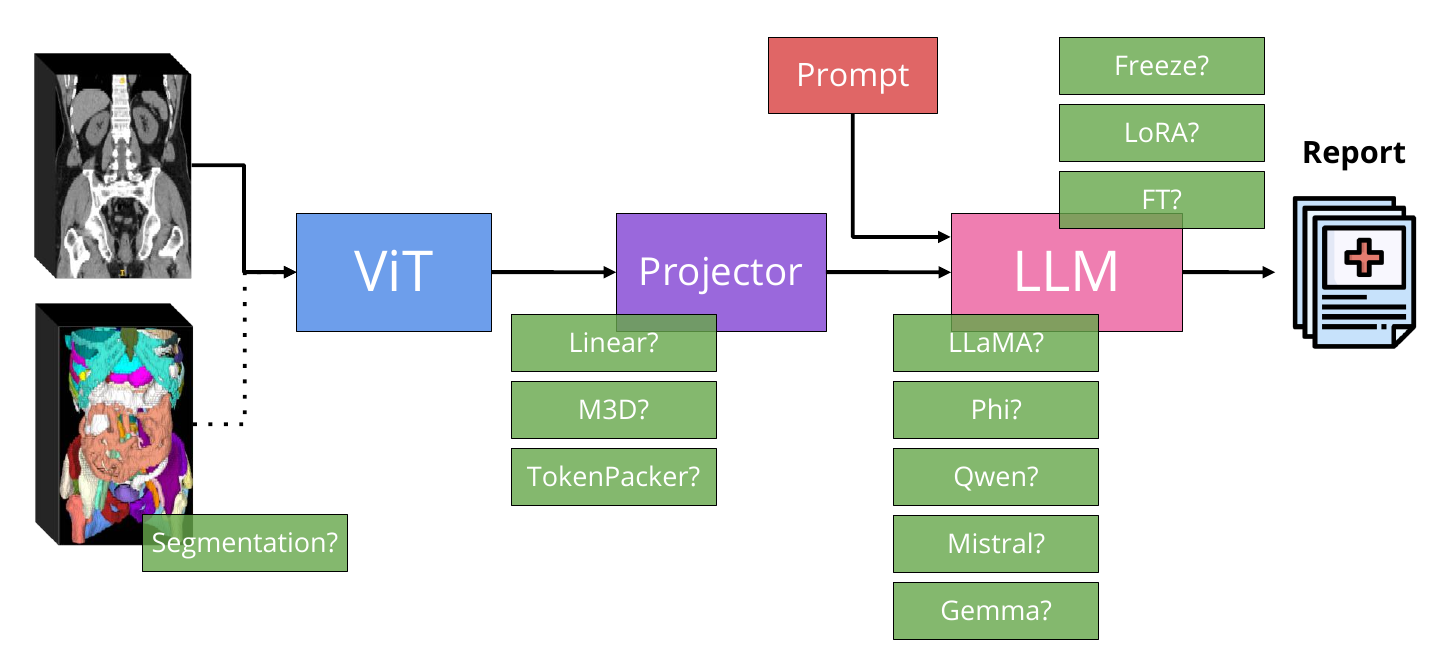}
    \caption{The design space of 3D MLLMs. We explore different choices in the selection of the visual representation, projectors, LLMs, and fine-tuning methods.}
    \label{fig:data_stats}
\end{figure}

\subsection{Visual Representations of 3D Inputs} \label{sec:vis_inputs}
The major challenge to address in representing 3D volumes for ViTs is the large number of tokens, which leads to a significant computational burden. Different techniques have been developed to reduce token count \cite{hilt,m3d,radfm}. AnyResolution \cite{llavanext} has been proposed to embed large image sizes, which divides the high-resolution input into multiple smaller crops, followed by concatenating and compressing their embeddings. We adopt this technique for processing larger 3D CT volumes.

\subsection{Projector Variants}
\label{sec:projector}
A naive  MLP projector is used as the baseline to project the visual embeddings to the same dimension of LLM input tokens. 
Spatial pooling perceiver (SPP) projector is the projector proposed in M3D \cite{m3d}, which  reduces the number of tokens while maintaining the 3D spatial structure.
TokenPacker~\cite{tokenpacker} reduce the number of tokens by interpolating visual features to low resolution. We extended TokenPacker to take in 3D inputs with one image size by expanding the depth dimension.

\subsection{Large Language Models and Fine-tuing Methods}\label{sec:llm}
We experimented with the instruction-tuned version of Llama 3.1 8B \cite{llama3}, Phi3-mini \cite{phi3}, Phi3-medium \cite{phi3}, Qwen 2.5 3B \cite{qwen}, Gemma 2B \cite{gemma}, Mistral 7B \cite{mistral}, and M3D pre-trained LLMs \cite{m3d}.
In addition to freezing the LLMs, we evaluated different fine-tuning techniques: parameter-efficient fine-tuning (PEFT) techniques like LoRA \cite{lora} and DoRA \cite{dora}, and full fine-tuning.

\subsection{Knowledge-base Report Augmentation}\label{sec:report_completion}
\begin{figure}[h]
    \centering
    \includegraphics[width=0.95\linewidth]{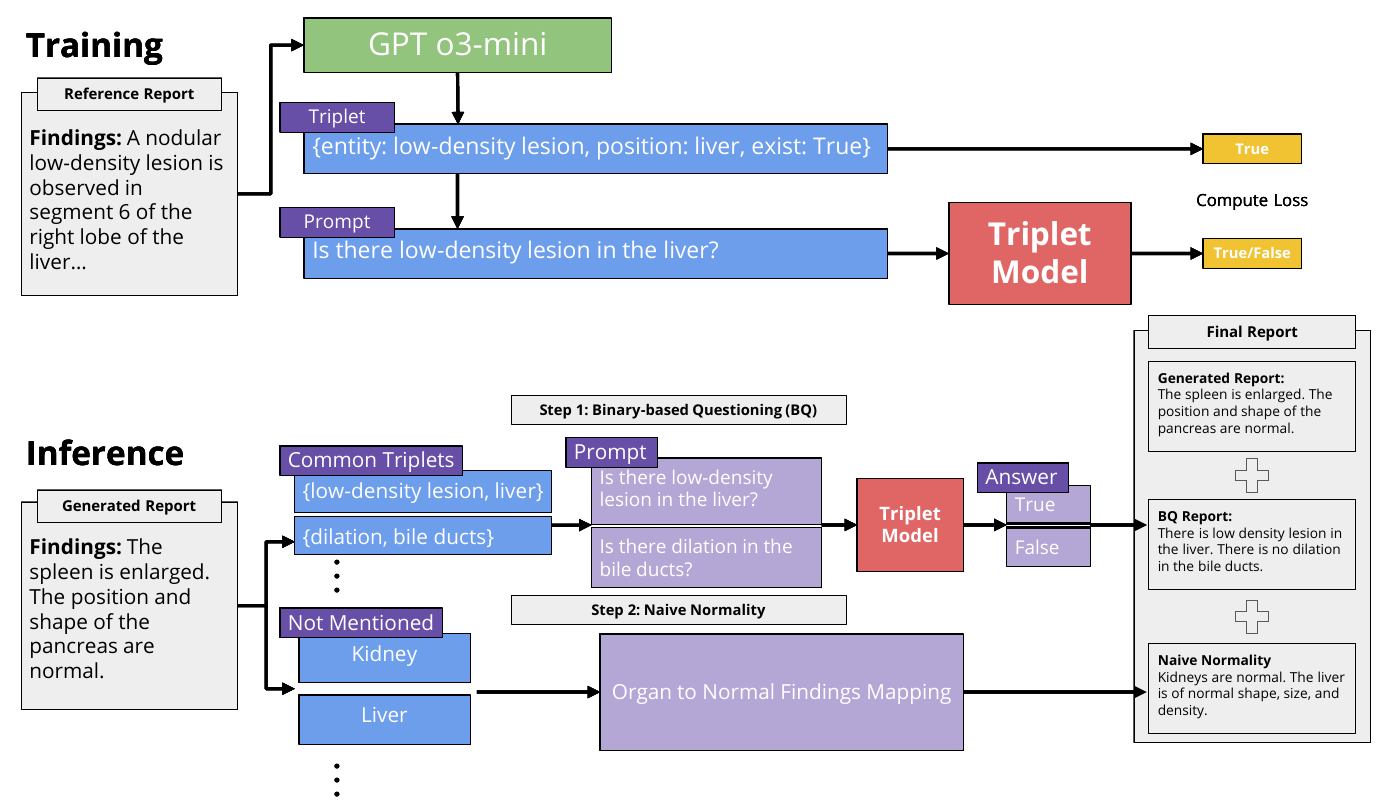}
    \caption{Knowledge-base report augmentation. After the MLLM generates a report, we apply two additional methods to augment the report, which are Binary-based Questioning (BQ) and naive normality. For BQ, we train a triplet model to answer questions about common conditions and use its answers to append additional findings to the generated report. To train the triplet model, we first prepare a set of question-answer pairs by prompting GPT o3-mini to generate triplets of the entity, location, and existence (in the format \{entity, location, exist\}) using findings from the reference reports. These triplets are then reformatted to questions, which are used to prompt the triplet model to generate binary True/False answers based on the CT volume. The value of "exist" is used as ground truth. At inference time, we prompt our triplet model with common triplets, which correspond to common findings, and based on its binary answer we append a positive or negative finding to the report. For naive normality, we append normal findings for organs that are not mentioned in the generated report after the BQ step. These two methods go hand-in-hand to ensure that our reports are complete and do not miss any common findings.}
    \label{fig:report_aug}
\end{figure}

We further explore ways to augment the generated reports to ensure completeness, using two introduced methods: Binary-based Questioning (BQ) and naive normality (NN) augmentation. Fig \ref{fig:report_aug} shows a graphical representation of the two methods.

\textbf{Binary-based Questioning (BQ).} In the first part of our report augmentation method, we make additional binary-based question inferencing for common findings on the three regions. Specifically, we first turn reports into triplets with the format \{entity, position, exist\} following \cite{medklip}. For example, the finding "A nodular low-density lesion is observed in the right lobe of the liver" becomes \{"low-density lesion", "liver", true\}. Then, we reformat the triplet into a question in the format "Is there \{entity\} in the \{position\}", with the "exist" being the ground truth. If no entity or position is detected, the question format becomes "Is the \{position\} normal?" or "Can you observe \{entity\} in this CT scan?", respectively.

The transformation into triplets was done by prompting GPT o3-mini. However, this may result in a lot of variations for the same finding statement (for example, \{enlargement, lymph nodes in the retroperitoneum\} and \{enlargement of lymph nodes, retroperitoneum\}). We then prompt GPT o3-mini again to go through all these variations and design a mapping that transforms them all into one common triplet. 

We train a model, which we will call the triplet model, to predict the \{exist\} variable for each triplet, where triplets are constructed from findings in the corresponding report. The triplet model is based on Phi3-mini with an M3D projector, trained for 200 epochs with the same hyperparameters described in Section \ref{sec:implementation}. Moreover, we collect the most common triplets and include them as questions in all examples throughout training. For these common triplets, if it, or any of its variations, is already in that CT's findings, we use the associated \{exist\} variable, otherwise, \{exist\} is set to false. This ensures that the model is always optimized for answering these common conditions. At inference time, we first use certain pre-defined keywords to ensure that the finding is not already mentioned in the generated report. Then, we prompt the model with the common triplets and map its binary answer to pre-defined findings. For example, an answer of "True" for the question "Are there nodules in the lung?" would be mapped to "Nodules are seen in the lungs." and "False" is instead mapped to "No nodules are seen in the lungs." These questions were designed to be general to improve performance on the metric, but this method can be adapted to specific triplets.

\textbf{Naive Normality (NN) augmentation.} In this method, we predefine a list of organs and conditions along with their associated normality finding for each of the three regions based on common findings from the dataset. Using this list, we then scan through the generated report and add a normality finding to organs or conditions not mentioned in the generated report. For example, in a chest report, we can scan through the generated report to find the word "heart." If it is not mentioned, we add the finding "The heart size and shape is normal and within limits. The heart is normal." to the report. Moreover, in normal findings that require two identifiers, like "No pleural effusion is seen in both pleural cavities or bilateral pleural cavities," we look for both the words "pleural effusion" and "pleural cavities" being in the same sentence.

\section{Experiments and Results}

\subsection{Dataset}
All the experiments are conducted using the MICCAI24 AMOS-MM challenge\footnote{MICCAI 2024 AMOS-MM challenge: \url{https://www.codabench.org/competitions/3137/}}, originally from the AMOS dataset \cite{amos}, which includes 1287 and 400 cases for training and validation. We used the validation set as our test set because the test set reports were not released. Each case contains one CT scan and the findings and impressions sections of the associated report for any or all the three regions: chest, abdomen, and pelvis. Chest findings are avaliable in 30.3\% of cases, abdomen for 99.8\%, and pelvis for 86.9\%. We follow the challenge setting and only use the findings as model output.
The dataset was gathered from Longgang District Central Hospital and Longgang District People's Hospital in Shenzhen, China.

\subsection{Implementation details}\label{sec:implementation}
A 3D ViT pre-trained on Radiopedia, from M3D \cite{m3d}, was used as the image encoder with a patch size of $4 \times 16^2$ and a volume size of $32 \times 256^2$. All MLLMs were trained for 150 epochs using a batch size of 4, with a final learning rate of $5e^{-5}$ and a cosine scheduler. We used a simple one-sentence prompt instructing the LLM to describe the findings in the CT scan. The Hugging Face Transformers \cite{transformers} framework was used for all LLM training and inferencing.

For processing larger CT volumes using AnyResolution \cite{llavanext}, the volume is divided into crops that are then embedded and concatenated before being passed to the LLM. This allows for handling high-resolution data with ViTs that are pre-trained on lower resolutions, while preserving intricate visual details. We use this method to process CT volumes of dimension 64 $\times$ $512^2$ using a ViT pre-trained on CTs of size 32 $\times$ $256^2$. In addition to the official metric in the competition, GREEN \cite{green}, we also computed the RaTEScore \cite{ratescore}, and commonly used text similarity metrics such as BLEU, ROUGE, and METEOR \cite{bleu,rouge,METEOR}.

\begin{table}[!htbp]
\centering
\caption{Effects of LLM, projector, and fine-tuning method on model performance.}
\makebox[\linewidth][c]{%
\begin{tabular}{l|c|c|cc|ccc}
\hline
& & & \multicolumn{2}{c|}{Clinical Metrics} 
      & \multicolumn{3}{c}{NLP Metrics} \\

LLM             & Projector         & FT Method 
                & GREEN & RaTEScore 
                & BLEU  & ROUGE & METEOR \\
\hline
Phi-3 mini 4B   & \multirow{7}{*}{M3D} & \multirow{7}{*}{Frozen}
                & 0.366 & \textbf{0.573} 
                & 0.272 & 0.384  & 0.357 \\
Phi-3 medium 14B&                      &
                & \textbf{0.370} & 0.572 
                & 0.269 & 0.382  & 0.356 \\
Gemma 2B        &                      &
                & 0.359 & 0.572 
                & 0.274 & 0.392  & \textbf{0.365} \\
Llama3.1 8B     &                      &
                & 0.359 & 0.570 
                & 0.264 & 0.379  & 0.355 \\
Qwen2.5 3B      &                      &
                & 0.341 & 0.554 
                & 0.252 & 0.378  & 0.342 \\
Mistralv0.3 7B  &                      &
                & 0.353 & 0.569 
                & 0.258 & 0.374  & 0.346 \\
M3D Phi-3       &                      &
                & 0.365 & 0.571 
                & \textbf{0.276} & \textbf{0.394} & 0.360 \\
\hline\hline
\multirow{3}{*}{Phi-3 mini}
    & M3D & \multirow{3}{*}{Frozen}
    & \textbf{0.366} & \textbf{0.573}
    & \textbf{0.272} & 0.384       & 0.357 \\
    & MLP &
    & 0.346          & 0.563
    & 0.268          & \textbf{0.386} & \textbf{0.359} \\
    & TP  &
    & 0.343          & 0.551
    & 0.259          & 0.372       & 0.349 \\
\hline\hline
\multirow{4}{*}{Phi-3 mini}
    & \multirow{4}{*}{M3D} & Frozen
    & \textbf{0.366} & \textbf{0.573}
    & \textbf{0.272} & \textbf{0.384} & \textbf{0.357} \\
    &                      & LoRA
    & 0.336          & 0.549
    & 0.253          & 0.361       & 0.343 \\
    &                      & DoRA
    & 0.321          & 0.546
    & 0.251          & 0.360       & 0.343 \\
    &                      & FT
    & 0.271          & 0.528
    & 0.232          & 0.341       & 0.323 \\
\hline
\end{tabular}%
}
\label{tab:llm_proj_ft_results}
\end{table}

\subsection{Results}
We report medical report generation results for different LLMs, projectors, and fine-tuning methods. We also explore different settings such as volume size, predicting impressions along with findings, and adding segmentation masks. We use the GREEN score \cite{green} as our base evaluation metric, but we also report RaTEScore \cite{ratescore} and text similarity metrics such as BLEU \cite{bleu}, ROUGE \cite{rouge} and METEOR \cite{METEOR}.

\textbf{LLMs.} First, we try multiple LLMs of different sizes ranging from 2B to 14B in size to gauge how much the LLM matters for this task. We show our findings in Table \ref{tab:llm_proj_ft_results}. Generally, we notice that the task is LLM-independent, for the LLMs we tried, and using a slightly larger LLM or better LLM does not significantly increase performance. However, we also notice that Qwen 2.5 performs worse than all the other models, which could be a result of its training data.

\textbf{Projector Variants.} We also benchmark different vision-to-text embedding projectors. Our results are shown in Table \ref{tab:llm_proj_ft_results}. We notice that the M3D projector \cite{m3d} outperforms both TokenPacker \cite{tokenpacker} and the naive MLP projector. This is likely due to the projector's ability to preserve 3D spatial information, compared to the other methods, which confirms the original work's findings \cite{m3d}.

\textbf{Fine-tuning Methods.} Moreover, we investigate the effects of using PEFT techniques like LoRA \cite{lora} and DoRA \cite{dora}, compared to full fine-tuning or keeping the LLM frozen. Table \ref{tab:llm_proj_ft_results} shows our results. We notice that, generally, increasing the number of tunable parameters in the LLM correlates negatively with performance, with a frozen LLM performing the best. This could be due to the task's sensitivity to overfitting or the relatively small scale of our dataset.

\subsubsection{Visual Input Representations.}
We experiment with increasing the image size and using the AnyResolution method \cite{llavanext}. For AnyResolution, we use a CT volume of size $64 \times 512^2$, and process crops of size $32 \times 256^2$, resulting in 8 non-overlapping crops being processed in total, per forward pass. We also pass in the original volume resized to $32 \times 256^2$, following the original implementation \cite{llavanext}. Table \ref{tab:vis_input} shows our results. Our results show that increasing the image resolution decreases the score across all metrics, on top of the increased computational burden. Going from the $32 \times 256^2$ image size to $32 \times 512^2$ increases the number of tokens for the ViT 4 times. This could be due to multiple factors, including the LLMs inability to handle longer contexts and unreliable evaluation metrics.

\begin{table}[!htbp]
\caption{Effect of increasing image resolution. Generally, we observe that increasing the image resolution decreases performance when the ViT is pre-trained on $32 \times 256^{2}$. Under similar compute, using AnyRes performs better than not.}
\label{tab:vis_input}
\centering
\begin{tabular}{cc|cc|ccc}
\hline
& & \multicolumn{2}{c|}{Clinical Metrics} & \multicolumn{3}{c}{NLP Metrics} \\
Vol.\ Size & AnyRes & GREEN & RaTEScore & BLEU & ROUGE & METEOR \\
\hline
$32 \times 256^{2}$ & F & \textbf{0.366} & \textbf{0.573} & \textbf{0.272} & \textbf{0.384} & \textbf{0.357} \\
$32 \times 512^{2}$ & F & 0.328 & 0.551 & 0.250 & 0.370 & 0.339 \\
$64 \times 512^{2}$ & T & 0.344 & 0.560 & 0.256 & 0.379 & 0.344 \\
\hline
\end{tabular}
\end{table}

\subsubsection{Additional Experiments.} We further experiment with several different training scenarios, such as predicting impressions along with findings and feeding a segmentation mask generated by TotalSegmentor \cite{totalsegmentator}, along with the CT volume, where we embed the volume and organ segmentation mask separately and concatenate their embeddings before being passed to the LLM. The results are shown in Table \ref{tab:other}. Predicting impressions slightly degrades performance. However, using the segmentation mask along with the volume slightly improves performance on both GREEN and RaTEScore, which could be because it is easier to identify abnormalities and organs with the help of the masks.

\begin{table}[!htbp]
\caption{Additional experiments. Predicting the impressions with findings slightly degrades performance, and using the segmentation mask improves performance. Only the findings were used in the evaluation.}
\label{tab:other}
\centering
\begin{tabular}{l|cc|ccc}
\hline
& \multicolumn{2}{c|}{Clinical Metrics} & \multicolumn{3}{c}{NLP Metrics} \\
Method & GREEN & RaTEScore & BLEU & ROUGE & METEOR \\
\hline
Baseline            & 0.366 & 0.573 & 0.272 & 0.384 & 0.357 \\
With Impressions    & 0.356 & 0.571 & \textbf{0.273} & 0.388 & \textbf{0.366} \\ 
With Segmentation   & \textbf{0.372} & \textbf{0.581} & 0.271 & \textbf{0.392} & 0.357 \\
\hline
\end{tabular}
\end{table}

\textbf{Knowledge-base Report Augmentation.} Lastly, we report the results for our introduced knowledge-base report augmentation methods. Table \ref{tab:report_aug} shows that both of our methods significantly increase the GREEN score, with naive normality increasing the average score by 8\%. We suspect that the disproportionately large increase in the pelvis GREEN is because the reports contain more normal findings. 
An important observation is that GREEN does not always correlate with the other metrics. Naive normality decreased the score text-similarity metrics and did not have an effect on RaTEScore, which could be indicative of the metric's robustness against these types of tricks.

\begin{table}[!htbp]
\caption{Knowledge-base report augmentation. Both methods significantly increase performance for GREEN. However, naive normality does not affect RaTEScore and decreases text-similarity metrics. Each row builds on the previous one: BQ is added to the baseline, and Naive Normality (NN) is applied on top of BQ. We report a P-value < 0.05 across all regions for GREEN and RaTEScore between baseline and the final reports (BQ + NN).}\label{tab:report_aug} 
\centering
\begin{tabular}{l|ccccc|ccc}
\hline
& \multicolumn{4}{c}{GREEN} & RaTEScore & BLEU & ROUGE & METEOR \\
Method                         & Chest & Abdomen & Pelvis & Avg.  & Avg.   & \multicolumn{3}{c}{Avg.} \\
\hline
Baseline                       & 0.243 & 0.358 & 0.499 & 0.366 & 0.573 & \textbf{0.272} & 0.384 & 0.357 \\
+ BQ             & 0.260 & 0.391 & 0.526 & 0.392 & 0.599 & 0.264 & \textbf{0.410} & \textbf{0.384} \\
+ NN & \textbf{0.287} & \textbf{0.415} & \textbf{0.708} & \textbf{0.470} & \textbf{0.601} & 0.207 & 0.390 & \textbf{0.384} \\
\hline
\end{tabular}
\end{table}

\begin{figure}[H]
    \centering
    \includegraphics[width=\linewidth]{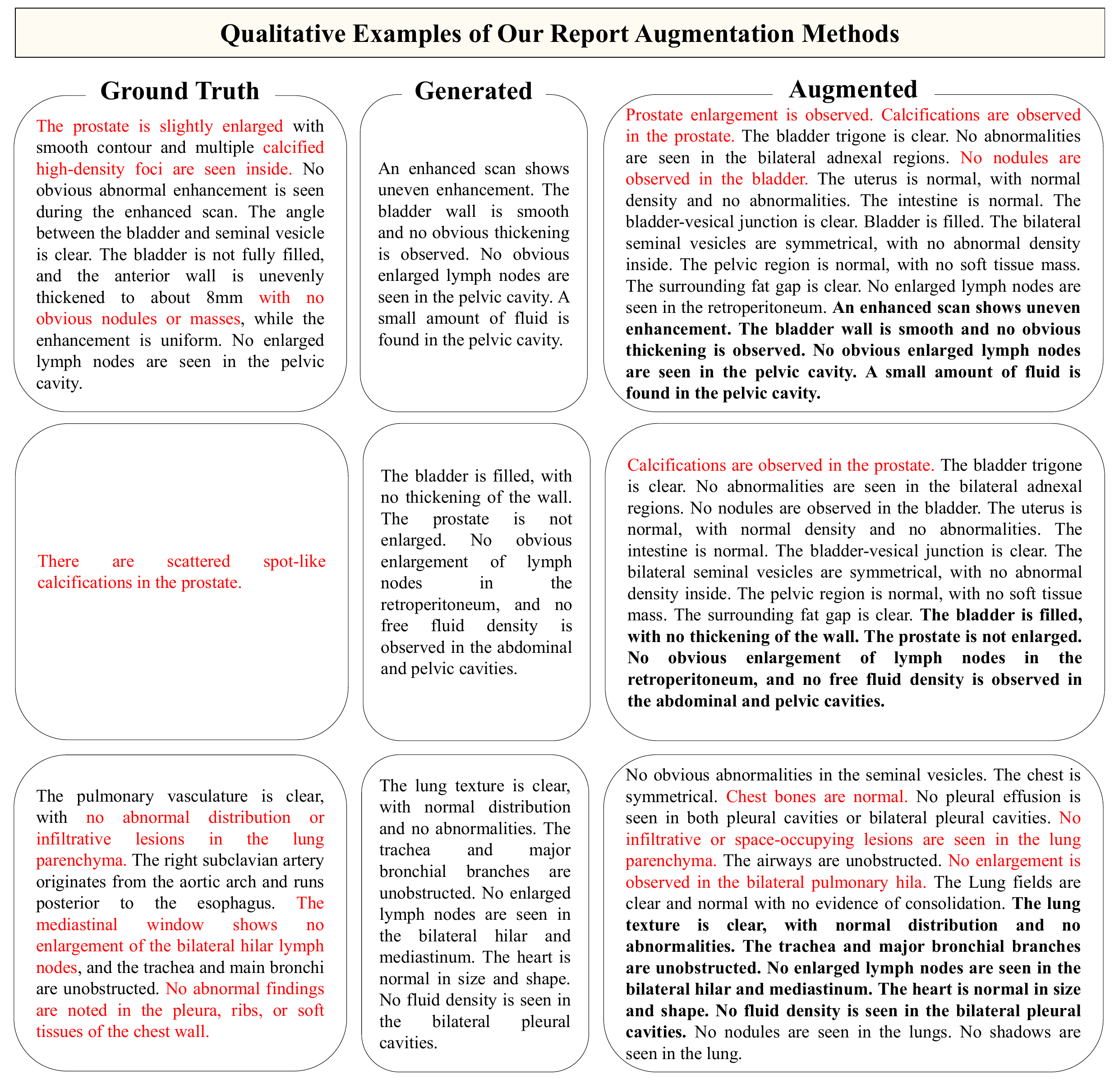}
    \caption{Qualitative examples of augmented reports. We show 3 examples of ground truth, generated, and the augmented reports after applying Binary-based Questioning and Naive Normality. Findings highlighted in red represents findings that were originally missed in the generated report, but that are then captured in the augmented one.Bolded findings in the augmented report represents the original generated report. All three examples show that our report augmentation methods are able to capture missed positive findings, while also ensuring completeness by mentioning normal organs explicitly.}
    \label{fig:qualitative_report_aug}
\end{figure}

\textbf{Qualitative Results.} We show in Figure \ref{fig:qualitative_report_aug} qualitative examples of the augmented reports after applying BQ and NN. Findings highlighted in red represents findings that are originally missed in the generated report, but are then captured in the augmented report. In the first report, the enlargement and calcification of the prostate are missed, but are then captured using additional inferences with our BQ triplet model. The added findings (For example, "Prostate enlargement is observed") are directly extracted from the knowledge-base, and this framework can be customizable based on the desired specificity of the findings. The last example shows an incomplete generated report that does not mention any observation about regions like the pleura and chest walls. These are then naively augmented using the NN method. Even though this does not change the inherent meaning of the report, it ensures that the final report is more explicit.

\section{Conclusion}
    
Our study reveals that CT report generation with 3D MLLMs, specifically on the AMOS-MM dataset, is largely LLM-independent, with models from 2B to 14B parameters showing similar performance. The M3D projector outperformed simpler methods by preserving 3D spatial structure, and freezing the LLM, compared to using parameter-efficient techniques or full fine-tuning. Moreover, increasing the input image resolution through the AnyResolution method did not enhance performance, underscoring the need for alignment between the pre-training and fine-tuning image resolutions. Lastly, our introduced knowledge-based report augmentation methods significantly boosted the GREEN score by up to 10\%. These insights contribute to a deeper understanding of the architectural choices for optimizing automated medical report generation, providing a robust foundation for future research in this domain.


\subsection*{Acknowledgement}
Our codebase is built upon the M3D \cite{m3d} GitHub repository. We also gracefully acknowledge Dr. Fan Bai for his invaluable suggestions and guidance. We also highly appreciate all the challenge organizers of the MICCAI 2024 AMOS-MM challenge for allowing us to use this great dataset.

\bibliographystyle{plain} 
\bibliography{mllm}

\begin{thebibliography}{10}

\bibitem{phi3}
Marah Abdin, Jyoti Aneja, Hany Awadalla, Ahmed Awadallah, Ammar~Ahmad Awan, et~al.
\newblock Phi-3 technical report: A highly capable language model locally on your phone.
\newblock {\em arXiv:2404.14219}, 2024.

\bibitem{m3d}
Fan Bai, Yuxin Du, Tiejun Huang, Max Q-H Meng, and Bo~Zhao.
\newblock M3d: Advancing 3d medical image analysis with multi-modal large language models.
\newblock {\em arXiv preprint arXiv:2404.00578}, 2024.

\bibitem{maira}
Shruthi Bannur, Kenza Bouzid, Daniel~C Castro, Anton Schwaighofer, Sam Bond-Taylor, Maximilian Ilse, Fernando P{\'e}rez-Garc{\'\i}a, Valentina Salvatelli, Harshita Sharma, Felix Meissen, et~al.
\newblock Maira-2: Grounded radiology report generation.
\newblock {\em arXiv:2406.04449}, 2024.

\bibitem{rise_ct}
R.~J.~M. Bruls and R.~M. Kwee.
\newblock Workload for radiologists during on-call hours: dramatic increase in the past 15 years.
\newblock {\em Insights into Imaging}, 11(1), 2020.

\bibitem{ViT}
Alexey Dosovitskiy, Lucas Beyer, Alexander Kolesnikov, Dirk Weissenborn, Xiaohua Zhai, Thomas Unterthiner, Mostafa Dehghani, Matthias Minderer, Georg Heigold, Sylvain Gelly, Jakob Uszkoreit, and Neil Houlsby.
\newblock An image is worth 16x16 words: Transformers for image recognition at scale.
\newblock In {\em International Conference on Learning Representations}, 2021.

\bibitem{llama3}
Abhimanyu Dubey, Abhinav Jauhri, Abhinav Pandey, Abhishek Kadian, Ahmad Al-Dahle, Aiesha Letman, Akhil Mathur, Alan Schelten, Amy Yang, Angela Fan, et~al.
\newblock The llama 3 herd of models.
\newblock {\em arXiv:2407.21783}, 2024.

\bibitem{ct}
C.~J Garvey.
\newblock Computed tomography in clinical practice.
\newblock {\em BMJ}, 324(7345):1077–1080, 2002.

\bibitem{ct2rep}
Ibrahim~E. Hamamci, Sezgin Er, and Bjoern Menze.
\newblock Ct2rep: Automated radiology report generation for 3d medical imaging.
\newblock In {\em International Conference on Medical Image Computing and Computer-Assisted Intervention}, pages 476--486, 2024.

\bibitem{lora}
Edward~J Hu, Yelong Shen, Phillip Wallis, Zeyuan Allen-Zhu, Yuanzhi Li, Shean Wang, Lu~Wang, and Weizhu Chen.
\newblock Lora: Low-rank adaptation of large language models.
\newblock {\em arXiv:2106.09685}, 2021.

\bibitem{amos}
Yuanfeng Ji, Haotian Bai, Chongjian Ge, Jie Yang, Ye~Zhu, Ruimao Zhang, Zhen Li, Lingyan Zhanng, Wanling Ma, Xiang Wan, et~al.
\newblock Amos: A large-scale abdominal multi-organ benchmark for versatile medical image segmentation.
\newblock {\em Advances in Neural Information Processing Systems}, 35:36722--36732, 2022.

\bibitem{mistral}
Albert~Qiaochu Jiang, Alexandre Sablayrolles, Arthur Mensch, Chris Bamford, Devendra~Singh Chaplot, Diego de~Las~Casas, Florian Bressand, Gianna Lengyel, Guillaume Lample, Lucile Saulnier, L'elio~Renard Lavaud, Marie-Anne Lachaux, Pierre Stock, Teven~Le Scao, Thibaut Lavril, Thomas Wang, Timoth{\'e}e Lacroix, and William~El Sayed.
\newblock Mistral 7b.
\newblock {\em arXiv preprint arXiv:2310.06825}, 2023.

\bibitem{METEOR}
Alon Lavie and Abhaya Agarwal.
\newblock Meteor: an automatic metric for mt evaluation with high levels of correlation with human judgments.
\newblock In {\em Proceedings of the Second Workshop on Statistical Machine Translation}, StatMT '07, page 228–231, USA, 2007. Association for Computational Linguistics.

\bibitem{cxr-llava}
Seowoo Lee, Jiwon Youn, Hyungjin Kim, Mansu Kim, and Soon~Ho Yoon.
\newblock Cxr-llava: a multimodal large language model for interpreting chest x-ray images.
\newblock {\em European Radiology}, January 2025.

\bibitem{llava-med}
Chunyuan Li, Cliff Wong, Sheng Zhang, Naoto Usuyama, Haotian Liu, Jianwei Yang, Tristan Naumann, Hoifung Poon, and Jianfeng Gao.
\newblock Llava-med: Training a large language-and-vision assistant for biomedicine in one day.
\newblock {\em Advances in Neural Information Processing Systems}, 36:28541--28564, 2023.

\bibitem{tokenpacker}
Wentong Li, Yuqian Yuan, Jian Liu, Dongqi Tang, Song Wang, Jianke Zhu, and Lei Zhang.
\newblock Tokenpacker: Efficient visual projector for multimodal llm.
\newblock {\em arXiv:2407.02392}, 2024.

\bibitem{rouge}
Chin-Yew Lin.
\newblock {ROUGE}: A package for automatic evaluation of summaries.
\newblock In {\em Text Summarization Branches Out}, pages 74--81, Barcelona, Spain, July 2004. Association for Computational Linguistics.

\bibitem{hilt}
Che Liu, Zhongwei Wan, Yuqi Wang, Hui Shen, Haozhe Wang, Kangyu Zheng, Mi~Zhang, and Rossella Arcucci.
\newblock Benchmarking and boosting radiology report generation for 3d high-resolution medical images.
\newblock {\em arXiv preprint arXiv:2406.07146}, 2024.

\bibitem{llavanext}
Haotian Liu, Chunyuan Li, Yuheng Li, Bo~Li, Yuanhan Zhang, Sheng Shen, and Yong~Jae Lee.
\newblock Llava-next: Improved reasoning, ocr, and world knowledge, 2024.

\bibitem{llava}
Haotian Liu, Chunyuan Li, Qingyang Wu, and Yong~Jae Lee.
\newblock Visual instruction tuning.
\newblock In {\em Advances in Neural Information Processing Systems}, 2023.

\bibitem{dora}
Shih-Yang Liu, Chien-Yi Wang, Hongxu Yin, Pavlo Molchanov, Yu-Chiang~Frank Wang, Kwang-Ting Cheng, and Min-Hung Chen.
\newblock Dora: Weight-decomposed low-rank adaptation.
\newblock {\em arXiv:2402.09353}, 2024.

\bibitem{green}
Sophie Ostmeier, Justin Xu, Zhihong Chen, Maya Varma, Louis Blankemeier, Christian Bluethgen, Arne Md, Michael Moseley, Curtis Langlotz, Akshay Chaudhari, et~al.
\newblock Green: Generative radiology report evaluation and error notation.
\newblock In {\em Findings of the Association for Computational Linguistics: EMNLP 2024}, pages 374--390, 2024.

\bibitem{bleu}
Kishore Papineni, Salim Roukos, Todd Ward, and Wei-Jing Zhu.
\newblock Bleu: a method for automatic evaluation of machine translation.
\newblock In {\em Proceedings of the 40th Annual Meeting on Association for Computational Linguistics}, ACL '02, page 311–318, USA, 2002. Association for Computational Linguistics.

\bibitem{gemma}
Gemma Team, Morgane Riviere, Shreya Pathak, Pier~Giuseppe Sessa, Cassidy Hardin, Surya Bhupatiraju, L{\'e}onard Hussenot, Thomas Mesnard, Bobak Shahriari, Alexandre Ram{\'e}, et~al.
\newblock Gemma 2: Improving open language models at a practical size.
\newblock {\em arXiv:2408.00118}, 2024.

\bibitem{work_load}
Christopher~J. Troupis, Richard A.~H. Knight, and Kenneth~K. Lau.
\newblock What is the appropriate measure of radiology workload: Study or image numbers?
\newblock {\em Journal of Medical Imaging and Radiation Oncology}, 68(5):530–539, 2024.

\bibitem{totalsegmentator}
Jakob Wasserthal, Hanns-Christian Breit, Manfred~T. Meyer, Maurice Pradella, Daniel Hinck, Alexander~W. Sauter, Tobias Heye, Daniel~T. Boll, Joshy Cyriac, Shan Yang, Michael Bach, and Martin Segeroth.
\newblock Totalsegmentator: Robust segmentation of 104 anatomic structures in ct images.
\newblock {\em Radiology: Artificial Intelligence}, 5(5), September 2023.

\bibitem{transformers}
T~Wolf.
\newblock Huggingface's transformers: State-of-the-art natural language processing.
\newblock {\em arXiv:1910.03771}, 2019.

\bibitem{medklip}
Chaoyi Wu, Xiaoman Zhang, Ya~Zhang, Yanfeng Wang, and Weidi Xie.
\newblock Medklip: Medical knowledge enhanced language-image pre-training in radiology.
\newblock {\em arXiv:2301.02228}, 2023.

\bibitem{radfm}
Chaoyi Wu, Xiaoman Zhang, Ya~Zhang, Yanfeng Wang, and Weidi Xie.
\newblock Towards generalist foundation model for radiology by leveraging web-scale 2d\&3d medical data.
\newblock {\em arXiv preprint arXiv:2308.02463}, 2023.

\bibitem{qwen}
An~Yang, Baosong Yang, Beichen Zhang, Binyuan Hui, Bo~Zheng, Bowen Yu, Chengyuan Li, Dayiheng Liu, Fei Huang, Haoran Wei, et~al.
\newblock Qwen2. 5 technical report.
\newblock {\em arXiv:2412.15115}, 2024.

\bibitem{ratescore}
Weike Zhao, Chaoyi Wu, Xiaoman Zhang, Ya~Zhang, Yanfeng Wang, and Weidi Xie.
\newblock Ratescore: A metric for radiology report generation.
\newblock {\em arXiv:2406.16845}, 2024.

\bibitem{medversa}
Hong-Yu Zhou, Subathra Adithan, Juli{\'a}n~Nicol{\'a}s Acosta, Eric~J Topol, and Pranav Rajpurkar.
\newblock A generalist learner for multifaceted medical image interpretation.
\newblock {\em arXiv preprint arXiv:2405.07988}, 2024.

\end{thebibliography}

\end{document}